# Multiscale Network Model for Fibrin Fibers and Fibrin Clot with Protofibril Binding Mechanics

**Sumith Yesudasan** [1, #a] **and Rodney D. Averett** [1*]

[1] School of Chemical, Materials, and Biomedical Engineering, University of Georgia, Athens, Georgia, United States of America

[#a] Current Address: Department of Mechanical Engineering, University of Jamestown, Jamestown, North Dakota, United States of America

[*]Corresponding author. Email: raverett@uga.edu

## Abstract.

The multiscale behavior of the individual fibrin fibers and fibrin clots is modeled by coupling atomistic simulation data and microscopic experimental data. We propose a protofibril element made up of nonlinear spring network, constructed based on the molecular simulation and atomic force microscopy results to simulate the force extension behavior of fibrin fibers. This new *network model* also accounts for the complex interaction of protofibrils with each other, effect of presence of solvent, Coulombic attraction and other binding forces. The network model is applied to simulate the force extension behavior of single fibrin fiber from atomic force microscopy experiments and shows good agreement. Thus validated fibrin fiber network model is then combined with a modified version of Arruda-Boyce eight chain model to estimate the force extension behavior of continuum level fibrin clot, which shows very good correlation. The results show that this network model is able to predict the behavior of fibrin fibers as well as fibrin clot at small strains, large strains and even closer to the break strain. We use the network model to explain why the fibrin clots and fibers doesn't behave like a worm like chain, instead behaves like a nonlinear spring.

## Introduction

Constitutive modeling of fibrin fibers and fibrin clots is still challenging and it is required to understand the origin of viscoelastic and hyper elastic type of behavior under various pathological states. Despite advancements in experimental methods to determine the molecular composition and crystal structure of fibrinogen, a major constituent of the blood clots, our understanding of blood clot mechanics remains poor, which is crucial in thrombolytic therapies and diagnosis of thrombosis [1]. In the past, researchers have utilized both the three-chain model and eight-chain model to determine the constitutive response [2] of fibrin clots. Although these models provide benefit for representing small-strain behavior, they have proven to be inadequate at fully representing the constitutive behavior of the clots at higher strains. In this paper, we present a simple nonlinear network model which is utilized to construct constitutive model of fibrin fiber and eventually fibrin clot.

The modeling of fibrin clot formation [3, 4] and fibrin mechanics [2, 5, 6] is of great importance in the scientific community [7]. Among them, the eight chain modeling of the fibrin fibers [2] and fibrin clots shows good agreement with the experimental results [8]. This model assumes the fibrin fibers as a combination of linear elastic component and a worm like chain (WLC) component, accounting for the fibrinogen unfolding in the macroscopic scale. However, the results from this model do not match well with the experiments at large strains. To address this issue, we model the protofibril element using a network model derived from the phenomenological modeling of fibrinogen from atomistic simulations and the D-E region interaction forces of rupture from atomic force microscopy (AFM) experiments. The proposed nonlinear network model was able to capture the force extension relationship of fibrin fibers at all strain levels. The validated fiber model is then combined with the modified version of Arruda-Boyce 8 chain model [9] and applied to a cylindrical shape fibrin clot stretching. The force extension results were matching well with experiments for small and large strain values.

**Multiscale construction of fibrin fibers**

The fibrin clot formation starts in the event of a thrombosis or hemostasis. The fibrinogen molecule is converted to fibrin monomers through the action of the enzyme thrombin. The fibrin monomers polymerizes predominantly through the knob-hole (A:a) interactions, and enhances the connectivity through γ-γ bonding, and weaker bonding through α-α and α-γ interactions. These interactions are stronger in the presence of Factor XIIIa and will form dimers, trimers, oligomers and eventually double stranded protofibrils which then cross links with each other forming fibrin fibers and network. Many detailed explanations related to the fibrin polymerization steps and molecular level details can be found in the literature [10-12].

A typical fibrin clot as seen from scanning electron microscopy (SEM) is shown in the Figure 1A. A simplistic schematic showing the physical construction of a single fibrin fiber in a clot is shown in the Figure 1B. The fibrinogen molecular model is shown in the bottom, double stranded protofibrils and sample inter connectivity shown in the middle and a matured fibrin fiber shown on the top. The construction of the protofibril strand, viewed from a mechanics perspective offers modeling it using mechanical spring network. The focus of this work is to simplify the bio network complex into a simple nonlinear network of mechanical springs, whose properties are estimated from atomic force microscopy (AFM) measurements or from molecular dynamics (MD) simulations.

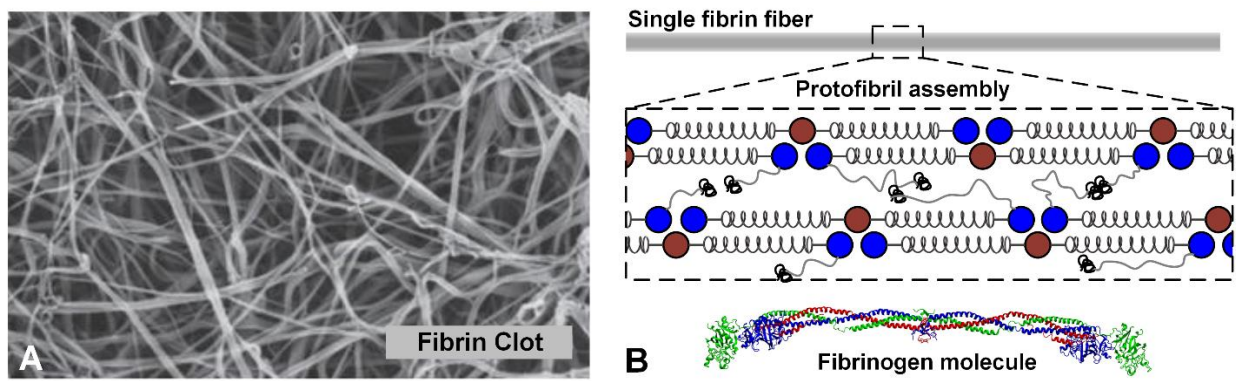

Figure 1: (A) Scanning electron microscopic image of human fibrin clot [13]. (B) Multiscale arrangement of a fibrin fiber. Hierarchical arrangement of fibrin (ogen) assembling to protofibrils, and their lateral aggregation to form fibers is shown here schematically.

**Modeling of the protofibril element**

The primary and most important component of the protofibril model will be the fibrin monomer or the fibrinogen molecule itself. The fibrinogen is a hexamer made up of pairs of three peptide chains α, β and γ. The crystal structure and molecular construction details of the fibrinogen molecule are available recently and is discussed elsewhere [14]. Consider the Figure 2, in which a portion of the long protofibril is shown on the top and a conceptual element shown in the bottom. The C-terminals of β and γ chains forms into nodules called D-region at either ends of the molecule, shown in blue colored circles in Figure 2. The N-terminals forms to be the central E-region. The fibrinogen molecule can be considered as a combination of two springs connected in series (Figure 2, sample molecule is 2-4-6 node sequence). The connection between the D-regions and the E-region of a fibrin monomer is modeled using a nonlinear force extension relationship graphically shown as red connection in Figure 2. The γ - γ interaction (3-5) is modeled through an amber colored connection as in Figure 2. In Figure 2, nodes 2, 3, 5 and 6 corresponds to D-region, 1, 4, 7 corresponds to E-region and nodes 8 and 9 corresponds to the dummy nodes.

The schematic of the protofibril element is shown as in Figure 2, which is assumed to be in quasi static equilibrium. The free body diagram shows the force balance of the element with the $F$ being force induced per monomer. In order to derive the force extension relation of the element, we need to understand the individual mechanical properties of its constituents. The force extension relationship of single fibrinogen molecules is available from both Atomic Force Microscopy (AFM) experiments [8, 15] and also from Molecular Dynamics (MD) simulations [8, 15, 16]. Often, the AFM results for fibrinogen force relationship doesn't show a consistent trend [8]. Hence, we will rely on the results from the MD simulations.

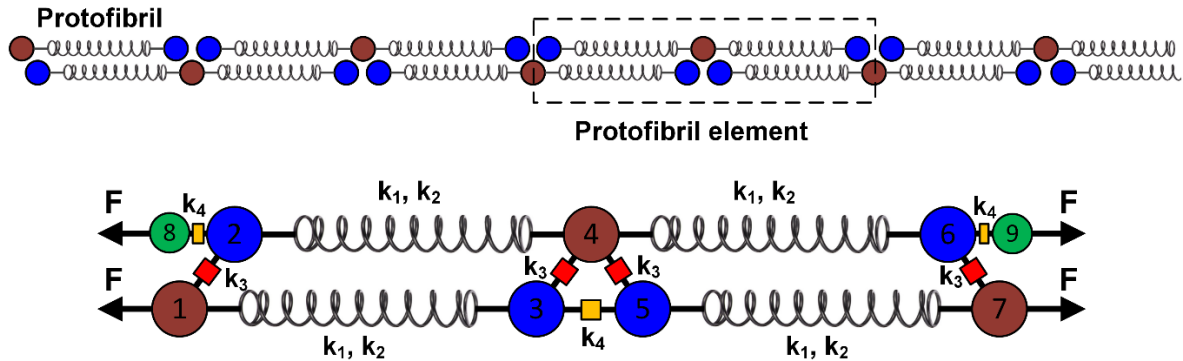

Figure 2: Development of a protofibril element using nonlinear spring network. A portion of the double stranded protofibril (top) is considered to develop the nonlinear protofibril element (bottom).

The force extension trend of fibrinogen molecule from MD simulations [8] shows a nonlinear response. This nonlinear force extension relation can be thought of a combination of linear spring and a cubic nonlinear spring as shown in the Eqn. 1.

$$F_{fibrinogen} = k_1 x + k_2 \text{H}[x - d_2](x - d_2)^3 \qquad (1)$$

Here, $k_1$ is the linear spring constant, $k_2$ is the nonlinear spring constant, $x$ is the extension, $d_2$ is the distance at which nonlinearity kicks in and H is the Heaviside function. The resulting profile is shown in Figure 3A, with a correlation coefficient of 0.96. The parameters took the values as shown in Table 1, for achieving this fit. The parameter $d_2$ is taken as 46 nm, the length of the fibrinogen, in order to relate the unfolding of the β-structure which occurs around 150-300 pN [17, 18], $k_1$ and $k_2$ are adjusted to match the results.

The interaction of D-region with E-region is considered to be relatively weak hydrophobic attraction and called as knob-hole interaction (A:a), modeled using an error function as shown in Eqn. 2. From literature, the force required to break one of these interactions ($k_3$) is highly likely around 130-140 pN [18, 19]. The Eqn. 2 shows the distance dependent force between D and E regions. A smaller value of $d_3$ will reduce the spread of the function, meaning tighter covalent like bonding and a larger value simulates plastic like behavior. Since no experimental data is available to estimate the flexibility of A:a interactions, we assume that the maximum elongation D-E coupling undergoes before rupturing to be 10 nm ($d_3$).

$$F_{D-E} = k_3 erf(x/d_3) \tag{2}$$

The interaction between γ and γ of the D regions is considered to be weak compared with the D-E interaction and is modeled using Eqn. 3. Since we couldn't find any quantified experimental data related to the rupturing force of D-dimers, we assume that it is in the similar order of magnitude of B:b knob:hole interactions [20]. This makes $k_4$ to be 10 times weaker than $k_3$ and $d_4$ twice larger than $d_3$.

$$F_{D-D} = k_4 erf(x/d_4) \tag{3}$$

The Eqn. 2 and Eqn. 3 are shown graphically in Figure 3B, which shows an initial linear like resistance and later leads to a plastic like elongation.

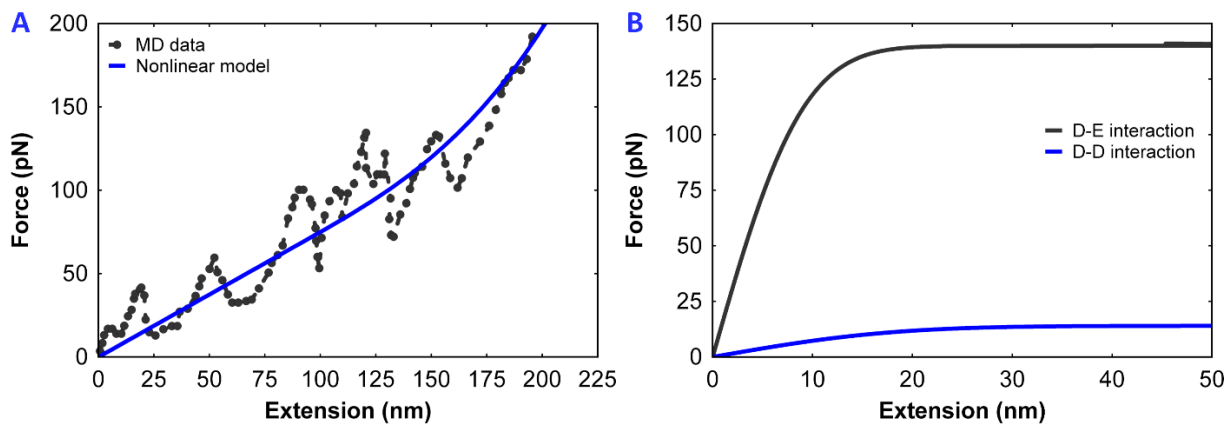

Figure 3: (A) Force extension comparison of a single fibrinogen molecule from Molecular Dynamics simulation data [8] and from our nonlinear model. (B) Force extension curve to simulate gamma-gamma interaction and A:a hole:knob interactions and their detachment. (C) Comparison of AFM data of a single fiber with our network model.

**Solution procedure**

The Figure 2 shows the protofibril element with nodes numbered from 1 to 7. For every node, we perform force balance conditions under quasi static loading assumption and derive the system of nonlinear force-extension equations. The equations can be found in the Appendix. This process is extended to the entire length of the protofibril. For reference and comparison, we used the similar parameters as in Liu et al.'s paper [21]. The length of the fiber is 12 $\mu m$, diameter as 330 $nm$, 30 % protein content on fiber and fibrin monomer radius of 2.25 $nm$. This gives 1614 monomers per cross sectional area and 261 protofibril elements along the length. Let us first consider the protofibrils are all parallel and not interconnected. This leads to the fibrin fiber force of $F_{fiber} = 1614F$, where $F$ is the force per fibrin monomer.

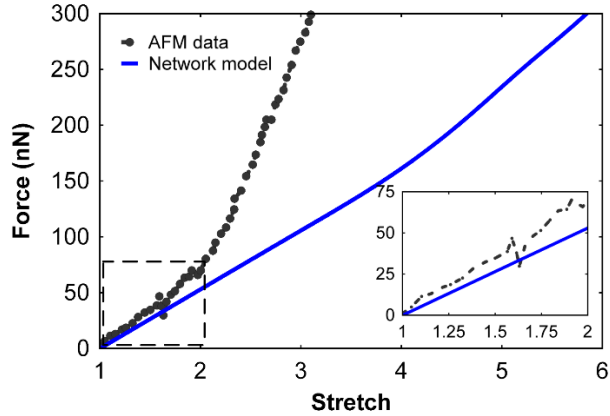

Figure 4: The force stretch relationship comparison of a 12 μm long 330 nm thick fibrin fiber with AFM experiments [17] (dotted line with markers) and with our network model (blue solid line). The inset shows the zoomed in location of the curve with almost close agreement initially.

The system of nonlinear equations thus deduced can be considered as a nonlinear minimization problem. The Jacobian for this system is sparse and singular which makes it difficult to be solved using standard methods like Levenberg – Marquardt algorithm [22]. We tried with other standard implementations in MATLAB software [23] and Trust Region Dogleg algorithm [24] gave good stability. The force $F$ is applied at 500 steps, starting from zero reaching to a maximum of 200 pN. The resulting force extension relationship is shown in Figure 4, comparing with the results from the AFM experimental data [21]. The comparison shows that they closely follow in the initial stage (inset of Figure 4) and later they do not match showing a less stiff fiber network model. This is not surprising, since at this stage we have not considered the effects of inter protofibril attraction, αC chain cross linking effects and effect of presence of solvent water etc.

## Results and discussion

It is important to include the effect of inter protofibril lateral interactions to account for the accurate mechanical properties of fibrin fibers. The primary contributors to this interaction is αC chain cross linking, binding of Factor XIIIa, presence of water around and inside fibrin fibers, and inter protofibril hydrophobic and Coulombic attraction. This makes the system complex and calls for atomistic details of these individual interactions which is unavailable in the literature. However, we can circumvent this problem by considering a unified force of binding between the protofibrils which participate in the event of force application. An ideal choice for this type of binding force is similar to the nonlinear model that we used for fibrinogen molecule. This includes a linear component and a distance dependent cubic nonlinear component. The nonlinear spring constant is tied to the linear constant by a factor to reduce the number of unknowns. The resulting equation is shown in Eqn. 4.

$$F_{Binding} = k_5 x_{elem} + k_6 H[x_{elem} - d_5](x_{elem} - d_5)^3 \qquad (4)$$

Here, $k_5$ is the linear stiffness of the spring, $k_6$ is the nonlinear stiffness, $x_{elem}$ is the elongation of a single protofibril element, $d_5$ is the distance at nonlinear spring takes effect. Based on our simulations, the parameters are found to be $k_5 = 0.1\ pN/nm$, $k_6 = \frac{k_5}{500} pN/nm^3$, and $d_5 =$ $\frac{L_{elem}}{4} nm$. Using these parameters, Eqn. 4 is connected across the protofibril element acting in parallel with the rest of the network. A value of zero for $k_5$ will switch off the effect of this binding force in the element. The resulting force extension curve of the network model is compared with the AFM experimental data in Figure 5. The results show very good agreement with the AFM data even up to the break strain limit.

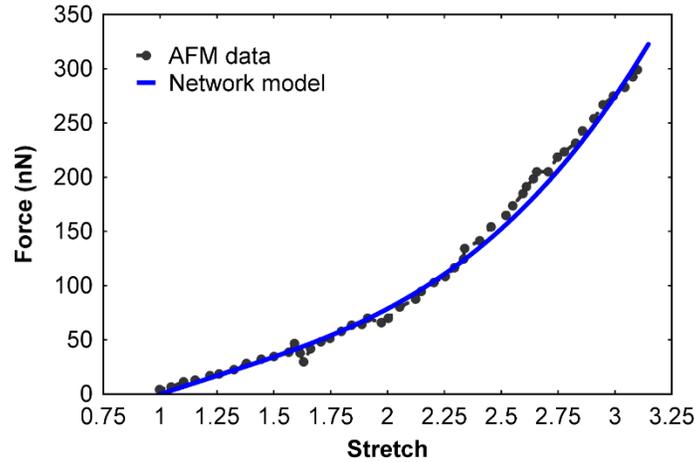

Figure 5: Fibrin fiber force-stretch relatiosnhip of our network model compared with Atomic Force Microscopy data [21]. The network model accounts for inter-protofibril binding forces.

The parameters used for various equations used in this work is summarized in the Table 1.

**Table 1:** Parameters and its values used in the study

| Parameter | Unit | Value | Parameter | Unit | Value |
|---|---|---|---|---|---|
| $k_1$ | $pN/nm$ | 1.5 | $d_2$ | $nm$ | 46 |
| $k_2$ | $pN/nm^3$ | 0.0003 | $d_3$ | $nm$ | 10 |
| $k_3$ | $pN$ | 130 | $d_4$ | $nm$ | 20 |
| $k_4$ | $pN$ | 13 | $d_5$ | $nm$ | 11.5 |
| $k_5$ | $pN/nm$ | 0.1 | | | |
| $k_6$ | $pN/nm^3$ | 2e-4 | | | |

**Sensitivity of the force extension curve**

To test the model sensitivity with radius and length of the fibers, we have changed the radius of the fibrin fiber from 25 nm to 225 nm by keeping the length as 12 μm. Next, we kept the radius constant as 165 nm and varied the length of the fibrin fiber from 0.5 μm to 14 μm. The resulting force extension curves are shown in Figure 6A and B. The trend of the curves show steeper slopes for increasing diameter of the fibers (Figure 6A) and steeper curves for decreasing length of the fibers (Figure 6B).

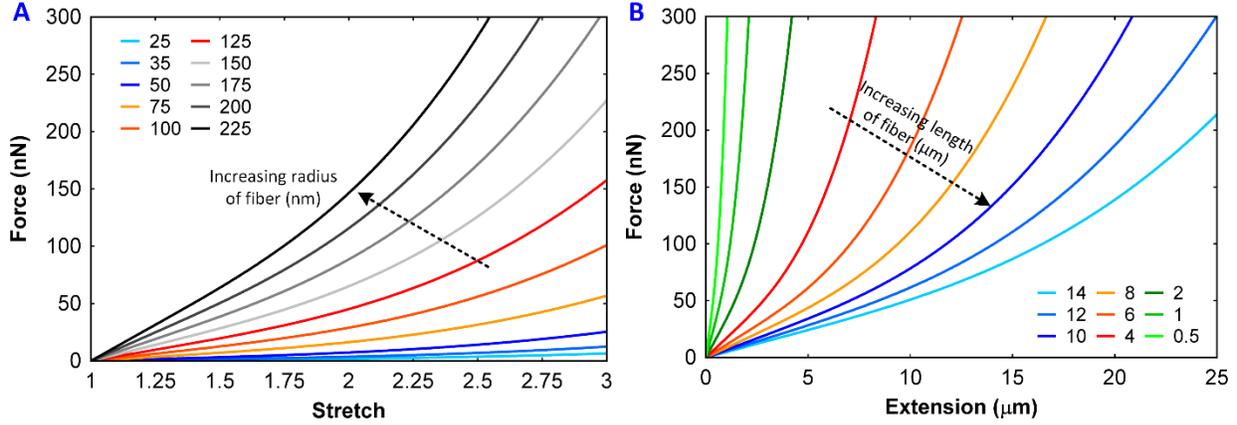

Figure 6: Force extension curves of fibrin fiber and its sensitivity with (A) diameter of the fibrin fibers and with (B) length of the fibers.

**Comparison with Fibrin Clot Experiment**

The next step is to apply the mesoscale network model to a continuum level fibrin clot. Previously, Brown and et al [2] have performed stretching of a 2mm diameter fibrin clot and developed a constitutive model to explain its behavior. However, the model doesn't correlate with the experimental results at higher stretch values. We will extend our fiber network model to this fibrin clot system using eight chain model (Figure 8B), originally developed by Arruda and Boyce [9]. With the assumption of incompressibility, the force extension relationship of fibrin cylindrical clot under uniaxial extension will takes the form as shown in the Eqn. 7. If the principal stretch $\lambda_1$ is aligned to the axis of extension, then incompressibility condition leads to the following relation.

$$\lambda_2 = \lambda_3 = 1/\sqrt{\lambda_1} \tag{5}$$

From the eight chain model, this leads to the stretch of the internal fiber as,

$$\lambda_{fiber} = \sqrt{\frac{\lambda_1^2 + 2/\lambda_1}{3}} \tag{6}$$

This fiber stretch along with the force extension relationship deduced from the network model as explained in the previous sections will lead to the following relationship for fibrin clot.

$$F_{clot} = \left(\lambda_1 - \frac{1}{\lambda_1^2}\right) \frac{\pi D^2 \nu L_{fiber}}{24 \lambda_{fiber}} F_{fiber} \tag{7}$$

Here, $F_{fiber}$ is the force extension relationship of the fibrin fiber deduced from the from the network model. $\lambda_1$ is the principal stretch aligned with the axis of uniaxial extension, $D$ is the diameter of the fibrin clot, taken as 2 mm, $\nu$ is the fiber density taken as 0.5 $\mu m^{-3}$ and $L$ is the fibrin fiber length. The parameters in Eqn. 7 are taken as the same in the literature [2] and that for the $F_{fiber}$ is the same as in Table 1 to be consistent. The resulting fibrin clot extension data is compared with the experimental results and also with the previously available worm like chain (WLC) based model as shown in the Figure 7.

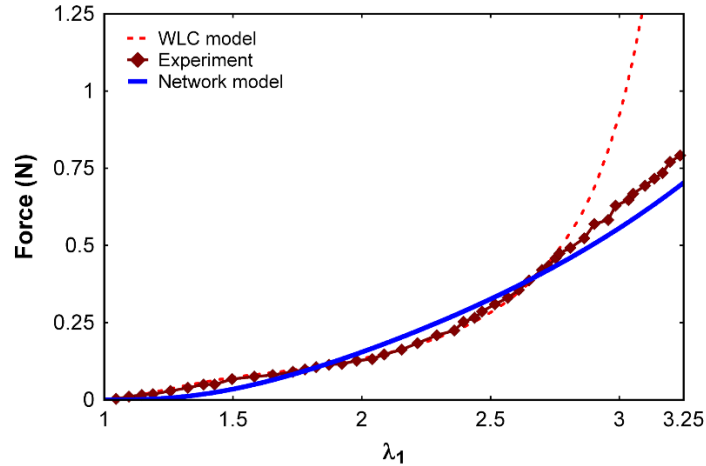

Figure 7: Comparison of force stretch relationship for fibrin clot using our network model with experimental data [2] and with worm like chain (WLC) model [2].

Results from Figure 7 shows that our network model closely captures the behavior of the clot mechanics seen in experiments. While the WLC based model deviates from the experiment at higher stretch values, our network model holds good. This is mainly due to the reason that fibrin fiber and clot behaves like a cubic like nonlinear material rather than behaving like a WLC. It is true that as a stiffer entity, DNA can be modeled as WLCs [25], but in the case of fibrinogen the AFM data [8] and molecular simulations [8] show a nonlinear behavior that doesn't fit well into a WLC model. Even from the experimental results, fibrin fiber (Figure 5) and fibrin clot (Figure 7),it doesn't look like a WLC trend, instead it shows a combination of both linear and nonlinear trend as suggested by our network model. In the case of WLC models, the chains after stretching closer to the maximum length, it becomes stiffer. But in the case of fibrin fibers, after stretching more than twice its original length, the force requirement increases for similar stretches, suggesting that it will be due to the unfolding of fibrinogen molecules and also the sliding of the protofibrils starts occurring leading to the nonlinear trend.

**Current assumptions and future improvements**

Currently, the model assumes that a single fiber is constructed from a cluster of network elements forming long protofibrils of length 12 $\mu m$ as shown in the Figure 8A. This is not the case in reality, where the individual protofibril is made up of 15-20 fibrin monomers [26] reaching a length around 0.5 $\mu m$. Splitting this into protofibril fragments will lead to the modeling of the entire fiber and the complex interconnectivity. The 8 chain model by Arruda-Boyce and its conceptual model is shown in Figure 8B. The fibrin fibers run from the corners of a rectangular volume towards the center and connects at the center. When the entire volume is subjected to external stretch or force, the internal fibers also get strained as shown in the figure. In the original 8 chain model, the authors use strain energy density function derived from the statistical behavior of the chains. In our work, we modify this and replace the fiber force extension relationship with our nonlinear network model.

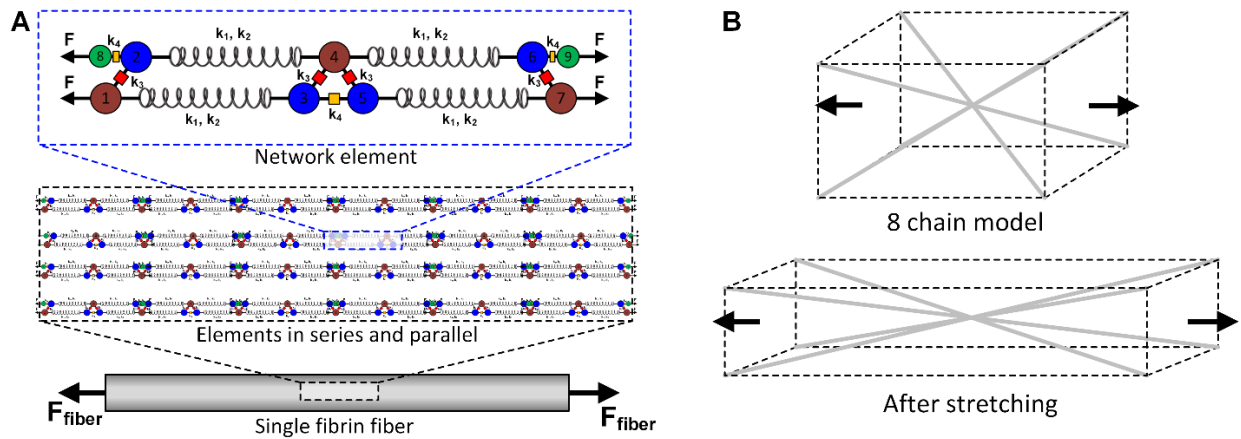

Figure 8: (A) Arrangement of network model element in series and parallel to form a fibrin fiber subjected to externally applied force. (B) Schematic of 8-chain model used for fibrin clot modeling. The light colored diagonally connecting lines represent the fibers.

## Acknowledgements

Research reported in this publication was supported by the National Heart, Lung, and Blood Institute of the National Institutes of Health under Award Number K01HL115486. The content is solely the responsibility of the authors and does not necessarily represent the official views of the National Institutes of Health. This study was also supported in part by resources and technical expertise from the Georgia Advanced Computing Resource Center, a partnership between the University of Georgia's Office of the Vice President for Research and Office of the Vice President for Information Technology.

## Conclusion

In this work, we have presented a simple yet accurate network model for the fibrin fibers and fibrin clots which is suitable to incorporate into standard finite element packages. The protofibril element is constructed based on the molecular simulation and atomic force microscopy results to simulate the force extension behavior of fibrin fibers. Also, when the interaction of protofibrils with each other, with solvent and other binding forces is included, then the force extension behavior of a single fibrin fiber matches exactly with the experiments. Thus validated fibrin fiber network model is then combined with the classical eight chain model to estimate the force extension behavior of continuum level fibrin clot, which shows very good correlation. The results show that this network model was able to predict the behavior of fibrin fibers as well as fibrin clot even when its extension was close to the break strain. We also explain the reason why the fibrin clots and fibers doesn't behave like a worm like chain as in the case of fibrinogen molecule using our network constitutive model.

## Appendix: Mathematical formulation of network element

The force balance of the protofibril element shown in Figure 2 at every node is given here.

$$N_1: k_3 erf\left(\frac{(x_2-x_1)}{d_3}\right) + k_1(x_3-x_1) + k_2 H(x_3-x_1-d_2)(x_3-x_1-d_2)^3 - F = 0 \tag{8}$$

$$N_2: k_1(x_4-x_2) + k_2 H(x_4-x_2-d_2)(x_4-x_2-d_2)^3 - 2k_3 erf\left(\frac{(x_2-x_8)}{d_3}\right) - k_3 erf\left(\frac{(x_2-x_1)}{d_3}\right) = 0 \tag{9}$$

$$N_3: k_4 erf\left(\frac{(x_5-x_3)}{d_4}\right) + k_3 erf\left(\frac{(x_4-x_3)}{d_3}\right) - k_1(x_3-x_1) - k_2 H(x_3-x_1-d_2)(x_3-x_1-d_2)^3 = 0 \tag{10}$$

$$N_4: k_1(x_6-x_4) + k_2 H(x_6-x_4-d_2)(x_6-x_4-d_2)^3 - k_1(x_4-x_2) + k_2 H(x_4-x_2-d_2)(x_4-x_2-d_2)^3 + k_3 erf\left(\frac{(x_5-x_4)}{d_3}\right) - k_3 erf\left(\frac{(x_4-x_3)}{d_3}\right) = 0 \tag{11}$$

$$N_5: k_1(x_7-x_5) + k_2 H(x_7-x_5-d_2)(x_7-x_5-d_2)^3 - k_3 erf\left(\frac{(x_5-x_4)}{d_3}\right) - k_4 erf\left(\frac{(x_5-x_3)}{d_4}\right) = 0 \tag{12}$$

$$N_6: k_3 erf\left(\frac{(x_7-x_6)}{d_3}\right) + 2k_3 erf\left(\frac{(x_9-x_6)}{d_3}\right) - k_1(x_6-x_4) - k_2 H(x_6-x_4-d_2)(x_6-x_4-d_2)^3 = 0 \tag{13}$$

$$N_7: F - k_3 erf\left(\frac{(x_7-x_6)}{d_3}\right) - k_1(x_7-x_5) - k_2 H(x_7-x_5-d_2)(x_7-x_5-d_2)^3 = 0 \tag{14}$$

$$N_8: 2k_3 erf\left(\frac{(x_2-x_8)}{d_3}\right) - F = 0 \tag{15}$$

$$N_9: F - 2k_3 erf\left(\frac{(x_9-x_6)}{d_3}\right) = 0 \tag{16}$$

The force value between the top end nodes (8th and 9th) should match and also the bottom end nodes (1st and 7th). However, the forces on bottom and top may not match due to the strength difference in the network. To obtain $F$ to be used in the $F_{fiber}$, we average them.